Zentropy theory for accurate prediction of free energy, volume, and thermal expansion without fitting parameters

Zi-Kui Liu, Nigel L. E. Hew, Shun-Li Shang

Department of Materials Science and Engineering, The Pennsylvania State University,

University Park, Pennsylvania 16802, USA

**Abstract:**

Based on statistical mechanics, a macroscopically homogeneous system, i.e., a single phase in the present context, is composed of many independent configurations that the system embraces. The macroscopical properties of the system are determined by the properties and statistical probabilities of those configurations with respect to external conditions.  The volume of a single phase is thus the weighted sum of the volumes of all configurations.  Consequently, the derivative of the volume to temperature of a single phase depends on both the derivatives of the volumes of every configuration to temperature and the derivatives of their statistical probabilities to temperature with the latter introducing non-linear emergent behaviors.  It is shown that the derivative of the volume to temperature of the single phase can be negative, i.e., negative thermal expansion (NTE), due to the symmetry-breaking non-ground-state configurations with smaller volumes than that of the ground-state configuration and the rapid increase of the statistical probabilities of the former, and NTE can be predicted without fitting parameters from the zentropy theory that combines quantum mechanics and statistical mechanics with the free energy of each configuration predicted from quantum mechanics and the partition function of each configuration calculated from its free energy.

# 1 Introduction

The stability criteria of a macroscopically homogeneous system require that the derivatives between conjugate variables in the combined law of thermodynamics must be positive except at the limit of stability where they become zero. However, there is no such requirement for derivatives between non-conjugate variables. The derivative of volume (V) to temperature (T) under constant pressure, i.e., thermal expansion, is thus not required to be positive in a stable system because V and T are not conjugate variables. Though the derivative of volume to temperature of a macroscopically homogeneous material, i.e., a single phase, is commonly thought to be positive, i.e., positive thermal expansion (PTE), it is known that the volume of water does decrease with the increase of temperature at temperature below 4°C, i.e., negative thermal expansion (NTE). The first manmade NTE material was an $Fe_{65}Ni_{35}$ alloy in 1897 by Guillaume [1] with near zero thermal expansion (ZTE) at room temperature, commonly referred as INVAR.

While there have been many theories that aim to describe NTE mechanisms in various materials, they are mostly phenomenological aiming for interpretation of experimental observations[2–9], and a fundamental understanding and a predictive theory without experimental inputs applicable to all materials are still lacking. Based on Maxwell relation, the derivative of V to T under constant pressure equals to the negative derivative of entropy to pressure under constant temperature. Therefore, the fundamental understanding of NTE is related to the entropy of the system. While the total entropy of a system can be accurately obtained as a function of temperature from integration of experimentally measured heat capacity, its pressure dependence is more difficult to comprehend, and its theoretical prediction remains elusive. In the present paper, the efforts by the author's group to accurately predict entropy of a single phase as a function of temperature and pressure are discussed in terms of the recently termed zentropy theory [10] along with its predictive capability of PTE, ZTE, and NTE without experimental inputs.

## 2  Statistical mechanics and entropy

Based on statistical mechanics, a single phase at finite temperature is composed of various independent configurations that are in statistical equilibrium with each other and its surroundings under given external constraints. The probability of each configuration is related to its and the system's partition functions as follows

$$p^k = \frac{Z^k}{Z} \qquad \text{Eq. 1}$$

where $p^k$ and $Z^k$ are the probability and partition function of configuration $k$, and $Z = \sum_{k=1}^{m} Z^k$ is the partition function of the system or the phase with $m$ being the number of independent configurations. The configurational entropy among the configuration is obtained as

$$S = -k_B \sum_{k=1}^{m} p^k \ln p^k \qquad \text{Eq. 2}$$

where $k_B$ is the Boltzmann constant.

For a closed system under hydrostatic pressure ($P$), the combined law of thermodynamics in terms of internal/total energy, $E$, is written as

$$dE = TdS - PdV \qquad \text{Eq. 3}$$

For a canonical ensemble, the combined law is written in terms of Helmholtz energy as follows

$$dF = -SdT - PdV \qquad Eq.\ 4$$

The partition functions of the system and its configurations are written as

$$Z = e^{-\frac{F}{k_BT}} = \sum_{k=1}^{m} e^{-\frac{E^k}{k_BT}} = \sum_{k=1}^{m} Z^k \qquad Eq.\ 5$$

where $E^k$ is the total energy of the configuration $k$. The Helmholtz energy of the phase can be obtained from above equation as follows

$$F = -k_BT lnZ = -k_BT \sum_{k=1}^{m} p^k lnZ - k_BT \left( \sum_{k=1}^{m} p^k lnZ^k - \sum_{k=1}^{m} p^k lnZ^k \right)$$
$$= \sum_{k=1}^{m} p^k E^k + k_BT \sum_{k=1}^{m} p^k lnp^k = \sum_{k=1}^{m} p^k E^k - TS \qquad Eq.\ 6$$

When there is only one configuration in the system, Eq. 6 becomes

$$F = E^k \qquad Eq.\ 7$$

Since $F = E^k - TS^k$ by definition, Eq. 7 gives $S^k = 0$ at finite temperature, indicating that the configurations are all pure quantum states without any unspecified internal degrees of freedom as implicitly implied by Gibbs as the quantum mechanics was not invented yet at that time and envisioned by Landau and Lifshitz [11]. For systems of practical interest, the number of pure quantum states is very large, and their complete sampling is in general intractable. The current available solution is their coarse graining through density functional theory (DFT) [12,13], resulting in a non-zero entropy for each configuration at finite temperature and thus the necessity to modify the formula of entropy and partition function in terms of the zentropy theory as discussed below.

## 3 Zentropy theory and DFT-based calculations

For configurations with non-zero entropy, it is necessary to add their contributions to the total entropy of the system. Our zentropy theory is schematically shown in Figure 1 with the following equation for the total entropy of the system [10,14,15]

$$S = \sum_{k=1}^{m} p^k S^k - k_B \sum_{k=1}^{m} p^k \ln p^k \qquad \text{Eq. 8}$$

The first summation in Eq. 8 reflects the bottom-up view of the system by considering the contribution from individual configurations, and the second summation represents the top-down view of the system, seeing the statistical fluctuations of configurations as discussed in Section 2 above. This nested formula of the zentropy theory can be extended to consider more complex systems such as forest, planet, and black holes with more degrees of freedom, with Eq. 8 representing one of the subsystems of the system [14,15]. This nested formula can also be extended in another direction to configurations with fewer degrees of freedom within the configuration $k$ until it reaches the ground-state configuration with its properties predicted by the density functional theory (DFT) [16]. The latter may provide some insights into superconducting and other interesting ground-state configurations as postulated by the present author [14,15].

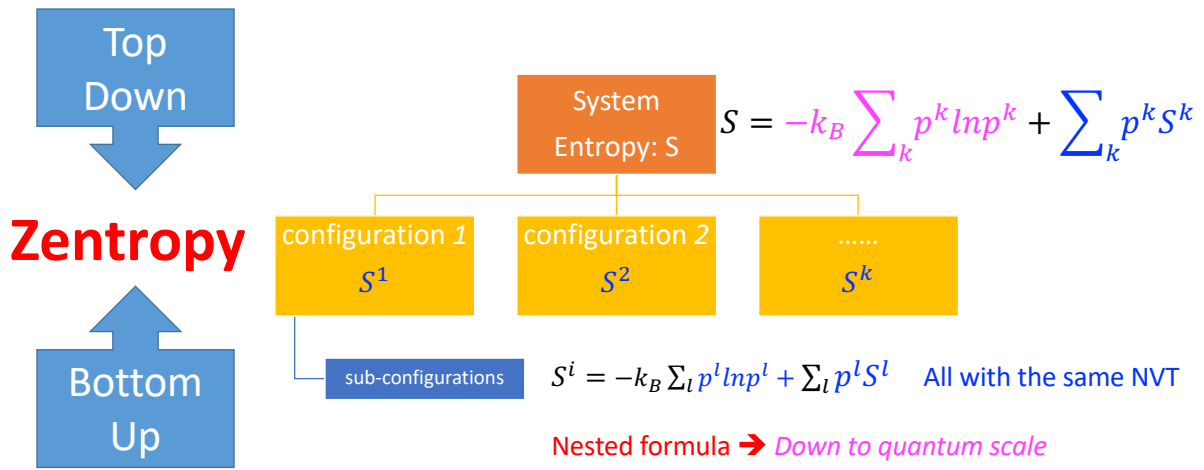

*Figure 1: Schematic top-down and bottom-up integration of the zentropy theory* [15].

*Reproduced with the permission of Ref. [15] Copyright © 2023, Elsevier.*

The Helmholtz energy of the system can thus be obtained as

$$F = \sum_{k=1}^{m} p^k E^k - TS = \sum_{k=1}^{m} p^k F^k + k_B T \sum_{k=1}^{m} p^k \ln p^k \qquad Eq.\ 9$$

where $F^k = E^k - TS^k$ is the Helmholtz energy of configuration $k$. Re-arranging Eq. 9 in the form of partition function, one obtains

$$Z = e^{-\frac{F}{k_B T}} = \sum_{k=1}^{m} e^{-\frac{F^k}{k_B T}} = \sum_{k=1}^{m} Z^k \qquad Eq.\ 10$$

$$p^k = \frac{Z^k}{Z} = e^{-\frac{F^k - F}{k_B T}} \qquad Eq.\ 11$$

Eq. 8 to Eq. 11 reduce to standard statistical mechanics when $S^k = 0$, i.e., pure quantum configurations with $F^k = E^k$ as discussed in Section 2.

Formulated as an exact theory of many-body systems, DFT [12] articulates that for an interacting electron gas there exists a universal functional of electron density such that the energy is at its minimum value, i.e., the ground-state energy with a unique ground-state electron density. The numerical solution is formulated by explicitly separating the independent-electron kinetic energy and long-range Coulomb interaction energy and replacing the many-body electron problem using independent valence electrons with an exchange-correlation functional of the electron density and an associated exchange-correlation energy and potential [13], i.e., *coarse graining of electrons*. Kohn and Sham [13] used the finite temperature generalization of ground-state energy of an interacting inhomogeneous electron gas by Mermin [17] and formulated the entropy of thermal electrons at finite temperatures. Wang et al [18] added the vibrational contribution and presented the Helmholtz energy as follows

$$F^k = E^{k,0} + F^{k,el} + F^{k,vib} = E^k - TS^k \qquad Eq.\ 12$$

$$E^k = E^{k,0} + E^{k,el} + E^{k,vib} \qquad Eq.\ 13$$

$$S^k = S^{k,el} + S^{k,vib} \qquad Eq.\ 14$$

where $F^{k,el}$, $E^{k,el}$, and $S^{k,el}$ are the contributions of thermal electron to Helmholtz energy, total energy, and entropy of configuration $k$ based on the Fermi–Dirac statistics for electrons, and $F^{k,vib}$, $E^{k,vib}$, and $S^{k,vib}$ are the vibrational contributions to Helmholtz energy, total energy, and entropy of configuration $k$ based on the Bose–Einstein statistics for phonons, respectively.

As the electron and phonon degrees of freedom are included in each configuration from the DFT-based calculations, all configurations are ergodic and symmetry-breaking in terms of magnetic spin, electric polarization, atomic short-rang ordering, and defects such as vacancy, dislocation, and stacking faults. For their $F^k$ to be predicted from DFT as a function of external stimuli, the non-ground-state configurations must be metastable. With their $p^k$ calculated from partition functions using $F^k$, the zentropy theory enables the integration of the quantum and statistical mechanics through Eq. 8 to Eq. 14 and is capable to predicting how a system responds macroscopically to external stimuli.

## 4   Derivative of volume to temperature

The pressure of a system can be calculated from the derivative of Helmholtz energy to volume as follows

$$P = -\frac{\partial F}{\partial V} = \frac{\partial(k_B T lnZ)}{\partial V} = \frac{k_B T}{Z}\frac{\partial(\sum_{k=1}^{m} Z^k)}{\partial V} = -\sum_{k=1}^{m} p^k \frac{\partial F^k}{\partial V} = \sum_{k=1}^{m} p^k P^k \qquad Eq.\ 15$$

where $P^k$ is the pressure of configuration $k$ evaluated at the system volume. In our previous publications as reviewed in refs. [14,15], the volume for given temperature and pressure was numerically evaluated from Eq. 15 along with the Helmholtz energy of the system. When the Helmholtz energy of the system is at its lowest value with one well, the system is in a single-phase region. When Helmholtz energy of the system with double or more wells can be lowered by separating into two phases with different volumes, the system is in a two or more phase region. This indicates that the zentropy theory can predict the Helmholtz energy of the system under metastable and unstable states, thus the free energy barrier between stable and metastable states. This is significant because the common wisdom is that the free energy of an unstable state could not be defined due to the imaginary vibrational modes that prevent the evaluation of its entropy. However, this view assumes that the atoms are static when evaluating the phonon properties, while the atoms are constantly moving at finite temperature. As all configurations used in the zentropy theory are stable, it does not depend on this unrealistic assumption in evaluating the free energy of unstable states.

As mentioned in our previous publications [10,19], we pointed out that one may consider the Gibbs ensemble, i.e., constant $N$, $P$, and $T$, in order to evaluate the derivative of volume to temperature under constant pressure. The combined law of thermodynamics is thus written in terms of Gibbs energy as follows,

$$dG = -SdT + VdP \qquad Eq.\ 16$$

From the Maxwell relation, one has

$$\frac{\partial V}{\partial T} = -\frac{\partial S}{\partial P} \qquad \text{Eq. 17}$$

The statistical mechanics in terms of Gibbs ensemble is shown below

$$Z = e^{-\frac{G}{k_B T}} = \sum_{k=1}^{m} e^{-\frac{G^k}{k_B T}} = \sum_{k=1}^{m} Z^k \qquad \text{Eq. 18}$$

$$G = \sum_{k=1}^{m} p^k G^k + k_B T \sum_{k=1}^{m} p^k \ln p^k \qquad \text{Eq. 19}$$

where $G^k$ is the Gibbs energy of configuration $K$. The volume and the derivative of volume to temperature is obtained as follows

$$V = \frac{\partial G}{\partial P} = -\frac{\partial (k_B T \ln Z)}{\partial P} = -\frac{k_B T}{Z} \frac{\partial (\sum_{k=1}^{m} Z^k)}{\partial P} = \sum_{k=1}^{m} p^k \frac{\partial G^k}{\partial P} = \sum_{k=1}^{m} p^k V^k$$
$$= V^g + \sum_{k=1}^{m} p^k (V^k - V^g) \qquad \text{Eq. 20}$$

$$\frac{\partial V}{\partial T} = \sum_{k=1}^{m} \left[ p^k \frac{\partial V^k}{\partial T} + \frac{\partial p^k}{\partial T} V^k \right] = \sum_{k=1}^{m} \left[ p^k \frac{\partial V^k}{\partial T} + \frac{\partial p^k}{\partial T} (V^k - V^g) \right] \qquad \text{Eq. 21}$$

where $V^g$ is the volume of the ground-state configuration. From Eq. 20, it can be seen that if $V^k < V^g$, it is possible $V < V^g$. With increase of temperature, the probability of the ground-state configuration decreases, i.e., $\frac{\partial p^g}{\partial T} < 0$, while the probabilities of non-ground-state configurations increases, i.e., $\frac{\partial p^{k \neq g}}{\partial T} > 0$.

The first term Eq. 21 is a linear combination of contributions from each configuration, and the second term gives nonlinear behavior. The condition for $\frac{\partial V}{\partial T} = 0$, i.e., ZTE, can be obtained as follows

$$\sum_{k=1}^{m} p^k \frac{\partial V^k}{\partial T} + \sum_{k=1}^{m} \frac{\partial p^k}{\partial T}(V^k - V^g) = 0 \qquad Eq.\ 22$$

As all variables in Eq. 22 are positive, and no contributions from $\frac{\partial p^g}{\partial T}$, the necessary condition to have a solution is for $V^k < V^g$, and the sufficient condition is that the volume decrease due to the second summation surpasses the weighted sum of the volume increase of individual configurations as shown by the first summation in the equation, i.e.,

$$\sum_{k=1}^{m} \frac{\partial p^k}{\partial T}(V^k - V^g) < -\sum_{k=1}^{m} p^k \frac{\partial V^k}{\partial T} \qquad Eq.\ 23$$

Both cases are shown in Figure 2 for Ce and Fe$_3$Pt, respectively, in the temperature-volume phase diagrams. For Ce, the volumes of antiferromagnetic (AFM) and ferromagnetic (FM) symmetry-breaking non-ground-state configurations are larger than that of the non-magnetic (NM) ground-state configuration, while for Fe$_3$Pt, the FM ground-state configuration has the largest volume. There is a critical point in both systems where the stable high temperature single phase becomes unstable and separate into two phases with the same crystal structure and different molar quantities such as volume, entropy, and magnetic spin configurations. The two phases at lower temperature and the stable single phase at high temperature are all composed of the same configurations, and the only difference among them is the probabilities of various configurations. As each configuration in all phases is stable, the instability of the macroscopically homogeneous single phase originates from the competition of various configurations when viewed from high temperature, rather than the conventional interpretation of phonon softening that considers phonon being stationary. While viewed from low temperature, the macroscopically homogeneous single-phase results from the mixture of two macroscopically homogeneous phases, and there is no instability involved.

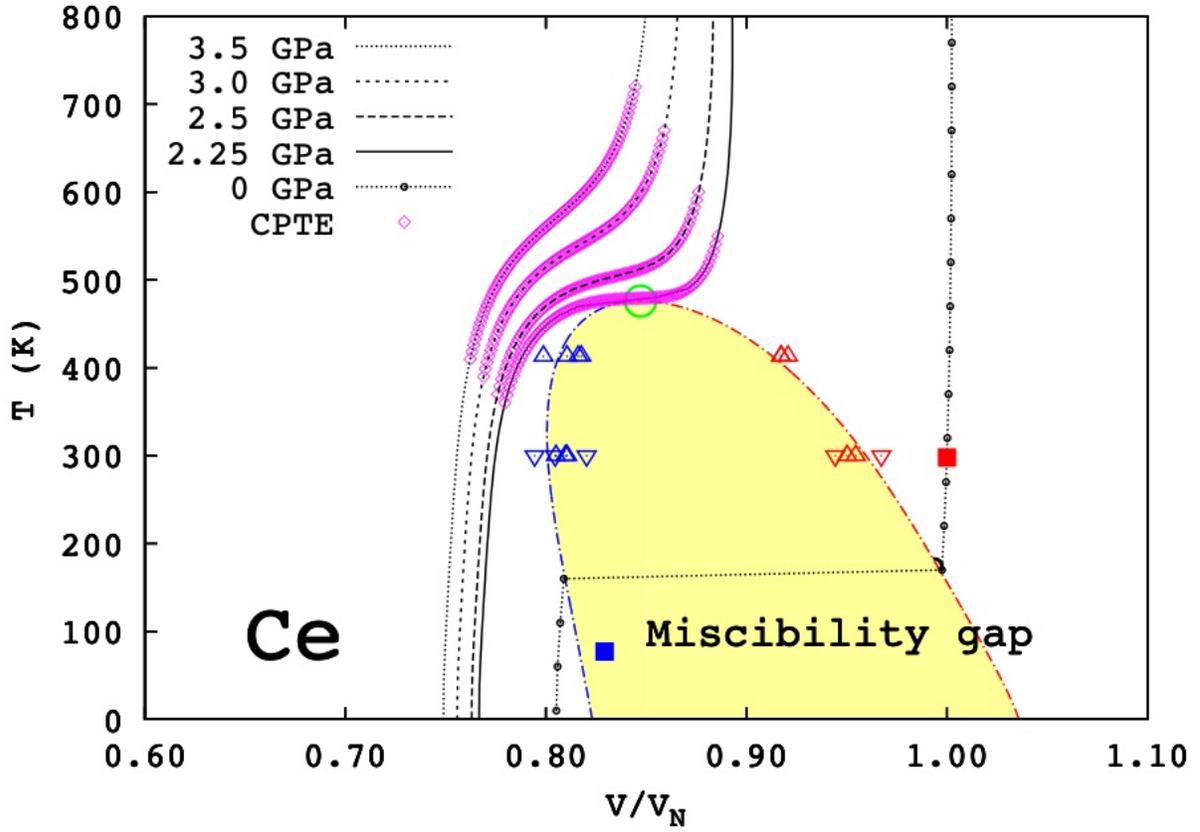

(a)

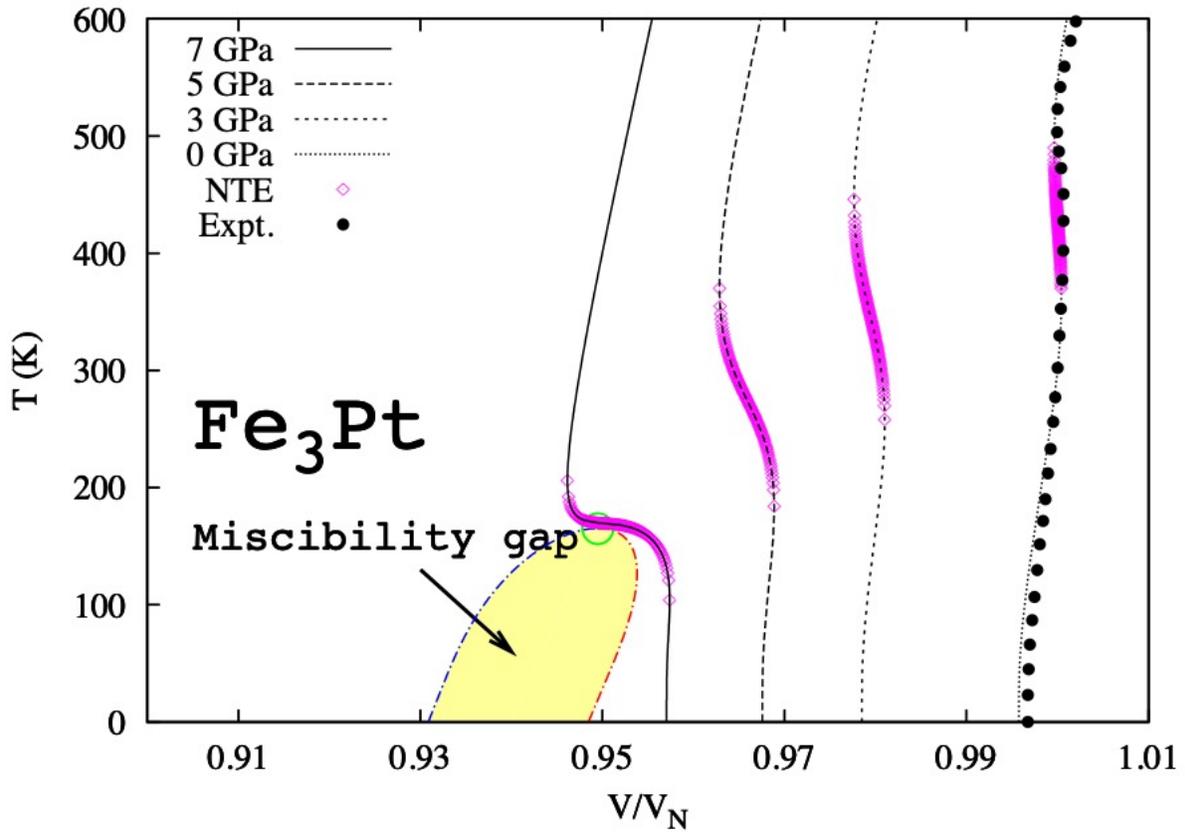

(b)

*Figure 2: Temperature-volume phase diagrams with isobaric volumes at various pressures of (a) Ce and (b) Fe$_3$Pt. The volume (V) is normalized to their respective equilibrium volume ($V_N$) at atmospheric pressure and room temperature. The highlighted regions by the pink open diamonds are illustrated for anomaly behaviors in terms of more positive thermal expansion for Ce and negative thermal expansion for Fe$_3$Pt. Below the critical point marked by the green open circle, the single phase is no longer stable and decomposes into a two-phase mixture in the region of a miscibility gap. Symbols are from various experimental measurements in the literature* [19]. *Reproduced with the permission of Ref.* [19] *Copyright © 2014, The Authors.*

## 5  Anharmonicity and emergent behaviors

Anharmonicity is usually represented by the deviation of entropy or heat capacity away from quasiharmonic behavior [5]. It is noted in Figure 2 that the volume change for $Fe_3Pt$ at a given pressure from 0 K to 600 K is rather small, supported by the experimental data as shown by the symbols on the isobaric volume curve under the ambient pressure. This indicates that the quasiharmonic approximation can give an accurate prediction of entropy of each configuration. From Eq. 8, it can be seen that the first summation is the linear combination of entropies of individual configurations, and the emergent behaviors, i.e., the behaviors that none of the individual configurations possess, originate from the second summation in the equation. This is the same for the derivative of volume to temperature shown by Eq. 21 where the emergent behavior of NTE is due to the rapid increase of the symmetry-breaking non-ground-state configurations and their volumes smaller than that of the ground-state configuration, and none of the individual configurations possess NTE.

For a stable system, the derivatives between a potential and its conjugate molar quantity in the combined law are positive, i.e.,

$$\frac{\partial T}{\partial S} > 0 \qquad \qquad Eq.\ 24$$

$$\frac{\partial (-P)}{\partial V} > 0 \qquad \qquad Eq.\ 25$$

When these derivatives become zero, the macroscopic system reaches its limit of stability and the extreme of anharmonicity, and the inverses of these derivatives diverge positively, i.e.,

$$\frac{\partial S}{\partial T} = \frac{\partial V}{\partial (-P)} = +\infty \qquad Eq.\ 26$$

Eq. 26 represents the heat capacity under constant pressure and can be derived from Eq. 8 as follows

$$\frac{C_P}{T} = \frac{\partial S}{\partial T} = \sum_{k=1}^{m} p^k \frac{\partial S^k}{\partial T} + \sum_{k=1}^{m} \left[ (S^k - S^g) + k_B \ln \frac{p^g}{p^k} \right] \frac{\partial p^k}{\partial T} \qquad Eq.\ 27$$

Again, the first summation is the linear combination of heat capacity of each configuration, and the nonlinear emergent behavior comes from the second summation.

However, derivatives between non-conjugate variables are not required to be positive, such as the derivative between volume and temperature. As they will also diverge at the limit of stability, they could be either positive or negative, i.e.,

$$\frac{\partial V}{\partial T} = \frac{\partial S}{\partial (-P)} = \pm\infty \qquad Eq.\ 28$$

As discussed above, the negative divergency occurs when the volume of the ground-state configuration is larger than those of non-ground-state configurations. The NTE spreads to single phase regions far away from the critical point as shown in Figure 2(b) with significant anharmonic behaviors.

More recently, the zentropy theory was applied to predict the ferroelectric-paraelectric (FE-PE) transitions in PbTiO$_3$ [20] with three configurations considered, i.e., tetragonal polarized configurations without domain wall (FEG), with 90° domain wall (90DW), and with 180° domain wall (180DW). With two-sets of domain wall energies predicted by DFT-based calculations at 0 K in the literature, the predicted FE-PE transition temperatures are 776 and 653 K, respectively, in comparison with experimental 763 K in the literature [20]. The present author's group is currently computing the free energies of the three configurations aiming for more accurate prediction. Our preliminary results on the energy-volume curves at 0 K are plotted in Figure 3, showing that equilibrium volumes at 0K are 603.79, 600.50, and 597.62 Å$^3$ for FEG, 90DW, and 180DW, respectively, in agreement with 603.42 and 599.88 Å$^3$ for FEG and 180DW configurations reported in the literature by the DFT-based calculations [21]. Based on the zentropy theory, the NTE in PbTiO$_3$ originates from the fact that the volume of the FEG ground-state configuration is larger than those of 90DW and 180DW symmetry-breaking non-ground-state configurations.

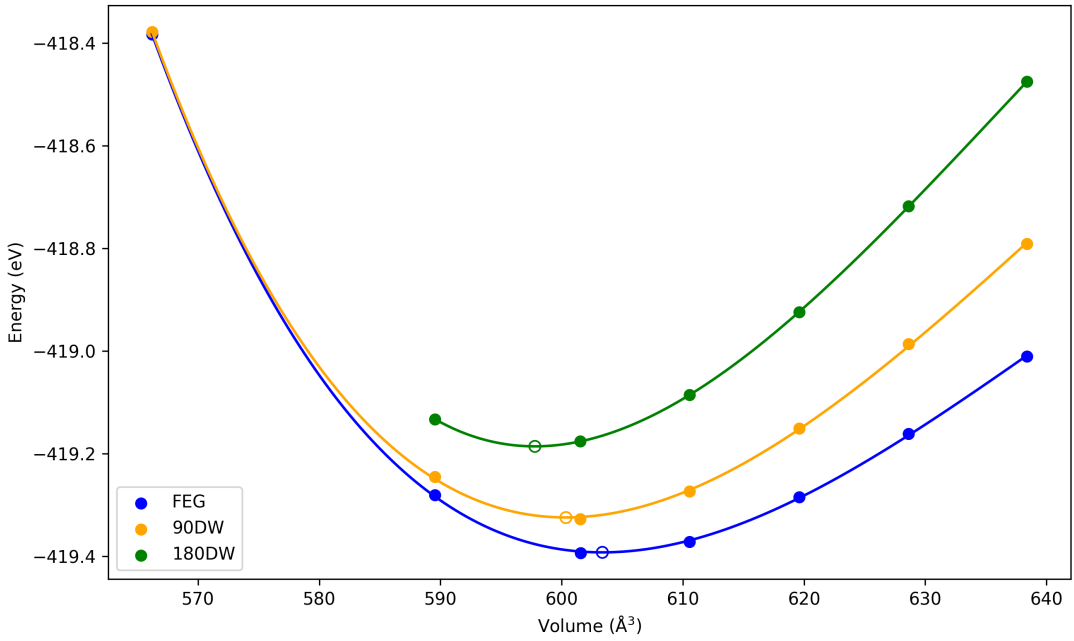

*Figure 3: Predicted energy-volume curves of PbTiO$_3$ at 0 K for FEG, 90DW, and 180DW configurations, respectively. The filled circles are from DFT-based calculations using the LDA exchange-correlation functional, while the curves were fitted using the third-order Birch*

*Murnaghan equation of states (EOS). The open circles represent the energy minimum from EOS fitting.*

# 6 Summary

The zentropy theory postulates that the entropy of a system contains contributions from entropies of the ground-state and symmetry-breaking non-ground-state configurations of a system and the statistical configurational entropy among these configurations. With the free energies of individual configurations predicted from DFT, the zentropy theory integrates quantum and statistical thermodynamics with the partition function of each configuration calculated from its free energy instead of total energy commonly used in the literature. With accurate free energy landscape of the system predicted by the zentropy theory, the properties of the system can be predicted by the first and higher-order derivatives of free energy with respect to its natural variables such as volume as the first derivative and thermal expansion as the second derivative. While the derivatives between conjugate variables are always positive for a stable system, the derivatives between non-conjugate variables such as thermal expansion can be either positive or negative with the latter due to the larger volume of the ground-state configuration than those of non-ground-state configurations in the system. It is articulated that the emergent behaviors and anharmonicity originate from the competition among the configurations, and their accurate predictions can be realized by the zentropy theory.

**Acknowledgements.** The present review article covers research outcomes supported by multiple funding agencies over multiple years with the most recent ones including the Endowed Dorothy Pate Enright Professorship at the Pennsylvania State University, U.S. Department of Energy (DOE)

Grant No. DE-SC0023185, AR0001435, DE-NE0008945, and DE-NE0009288, and U.S. National Science Foundation (NSF) Grant No. NSF-2229690.

**DECLARATIONS.** The manuscript was written by Liu and revise by Hew and Shang. Figures 1 and 2 were taken from the literature with references, and Figure 3 was calculated by Hew and Shang.